\documentclass[prb,twocolumn]{revtex4}
\usepackage{graphicx}

\newcommand{\R}{\mathbf{r}}

\newcommand{\UP}{n_{\uparrow}}
\newcommand{\DN}{n_{\downarrow}}

\newcommand{\be}{\begin{equation}}
\newcommand{\ee}{\end{equation}}
\newcommand{\bea}{\begin{eqnarray}}
\newcommand{\eea}{\end{eqnarray}}
\newcommand{\bean}{\begin{eqnarray*}}
\newcommand{\eean}{\end{eqnarray*}}

\newenvironment{acknowledgment}{{\flushleft \bf Acknowledgments:}}{}

\begin{document}

\title{ 
Laplacian-level density functionals for the\\
  kinetic energy density and exchange-correlation energy
}
\author{ John P. Perdew and Lucian A. $\mathrm{Constantin}^*$\\
\footnotesize{Department of Physics and Quantum Theory Group,}\\
\footnotesize{
Tulane University, New Orleans, LA 70118}}

\date{}

\begin{abstract}
We construct a Laplacian-level meta-generalized gradient approximation 
(meta-GGA) for the non-interacting (Kohn-Sham orbital) positive kinetic energy 
density $\tau$ of an electronic ground state of density $n$. 
This meta-GGA is designed to recover the  
fourth-order gradient expansion $\tau^{GE4}$ 
in the appropiate slowly-varying limit 
and the  von Weizs\"{a}cker expression $\tau^{W}=|\nabla n|^2/(8n)$ in the 
rapidly-varying 
limit. It is constrained to satisfy  
the rigorous lower bound 
$\tau^{W}(\mathbf{r})\leq\tau(\mathbf{r})$.
Our meta-GGA is typically 
a strong improvement over the gradient expansion of $\tau$
for atoms, spherical jellium clusters, 
jellium surfaces, the Airy gas, Hooke's atom, 
one-electron Gaussian density, quasi-two dimensional electron 
gas, and nonuniformly-scaled hydrogen atom. 
We also construct a Laplacian-level 
meta-GGA for exchange and correlation
by employing our approximate $\tau$ in the Tao, Perdew,
Staroverov and Scuseria 
(TPSS) meta-GGA 
density functional. The Laplacian-level TPSS 
gives almost the same 
exchange-correlation enhancement factors and energies as the full TPSS,
suggesting
that $\tau$ and $\nabla^2 n$ carry about the same information
beyond that carried by $n$ and $\nabla n$.
Our kinetic energy density integrates to an orbital-free kinetic energy
functional that is about as accurate as the fourth-order gradient expansion for
many real densities
(with
noticeable improvement in molecular atomization energies)
, but considerably more accurate for rapidly-varying ones.

\end{abstract}

\maketitle

\section{Introduction}
\label{sec1}
\noindent

In ground-state density functional theory, the noninteracting 
kinetic energy (KE) $T_{s}$ of a system of N electrons
 may be treated as an exact functional of 
the occupied Kohn-Sham (KS) \cite{KS} orbitals
$\{\phi_{i} \}$, and only the exchange-correlation energy has to be approximated. 
However finding an accurate orbital-free KE functional \cite{LK11} 
will simplify and speed up by orders of magnitude any KS self-consistent 
calculation.
In general, a kinetic energy density (KED) is any function 
which integrates to the non-interacting kinetic energy $T_{s}$:
\begin{equation}
T_{s}[\UP,\DN]=\int d\R\; \tau,
\label{e355}
\end{equation}
where $\UP(\R)$ and $\DN(\R)$ are the electron spin densities and
$\tau(\R)=\tau_\uparrow([\UP],\R)+\tau_\downarrow([\DN],\R)$.
Because of the local version of the spin-scaling relation \cite{a57}
\begin{equation}
\tau_\sigma([n_{\sigma}],\R)=(1/2)\tau([n=2n_{\sigma}],\R),
\label{e3551}
\end{equation}
we will only need to show our expressions for
spin-unpolarized systems with $\UP=\DN=n/2$.
There are two important forms of the KED: one which depends 
on the Laplacian of the Kohn-Sham orbitals and follows 
directly from the Kohn-Sham 
self-consistent equations:
\begin{equation}
\tau^{L}(\R)=-(1/2)\sum^{N}_{i=1}\phi^{*}_{i}(\R)\nabla^{2}\phi_{i}(\R),
\label{e356}
\end{equation}
and another which is positive definite:
\begin{equation}
\tau(\R)=(1/2)\sum^{N}_{i=1}|\nabla\phi_{i}(\R)|^{2}=
\tau^{L}(\R)+\frac{1}{4}\nabla^{2}n(\R),
\label{e357}
\end{equation}
where $n(\mathbf{r})=\sum_{i=1}^{N}|\phi_{i}(\mathbf{r})|^2$ 
is the electronic density. (We use atomic units, with 
$\hbar=m=e^{2}=1$). 
$(\tau^L+\tau)/2$ has been proposed \cite{LMM}
as the closest analog of a classical KED.

While a generalized gradient approximation (GGA) 
uses only $n$ and $\nabla n$, a meta-GGA (MGGA), such as
the accurate
nonempirical TPSS \cite{TPSS1} density functional for the exchange-correlation energy,
is constructed from local ingredients $n(\R)$, $\nabla n(\R)$, 
and $\tau(\R)$.
In this work, we will present evidence that $\tau(\R)$ and $\nabla^2 n(\R)$
can
carry essentially the same information
beyond that carried by $n(\R)$ and $\nabla n(\R)$.
To do so, and to make
an improved semi-local density functional for $T_s$, we shall construct a
Laplacian-level meta-GGA for $\tau(\R)$. 
Our semi-local 
functional recovers the fourth-order gradient expansion (GE4) KED 
for a slowly-varying density and the von  
Weizs\"{a}cker KED for a rapidly-varying density.

The gradient expansion (GE), which becomes exact for densities 
that vary slowly over space \cite{PCSB}, is \cite{YPKZ}
\begin{equation}
\tau = \tau^{(0)}F_s(p,q,...),
\label{grexp}
\end{equation}
with
\begin{equation}
F_s = \sum_{n=0}^{\infty} F^{(2n)}_s,
\label{grexp111}
\end{equation}
where $p=|\nabla n|^{2}/\{4(3\pi^{2})^{2/3}n^{8/3}\}$ and $q=\nabla^2 
 n/\{4(3\pi^{2})^{2/3}n^{5/3}\}$ are dimensionless
derivatives of the density, $\tau^{(0)}=\frac{3}{10}
(3\pi^{2})^{2/3}n^{5/3}$ is the Thomas-Fermi KED \cite{TF}, and 
$F^{(2n)}_{s}(p,q,...)$
is
the enhancement factor of the
2n-th term of the gradient expansion. The zero-th order  
enhancement factor is \cite{TF}
 $F^{(0)}_{s}=1$, 
and the second-order one is \cite{Ki,BJC}
$F_s^{(2)} = (5/27)p + (20/9)q$. The term linear in $q$
is a key ingredient in our MGGA.
Although this term integrates to zero in Eq. (\ref{e355}),
it is \cite{YPL} important for the KED. This term
also indicates rapidly-varying density regimes 
(e.g., near a nucleus where $q\rightarrow -\infty$ and in
the tail of the density where $q\rightarrow\infty$).
The fourth-order enhancement factor is \cite{Ho}
\begin{equation}
F^{(4)}_{s}=\Delta=(8/81)q^{2}-(1/9)pq+(8/243)p^{2}\geq 0.
\label{ef4}
\end{equation}
This enhancement factor
is a simplified expression obtained with Green's theorem 
(integration by parts in Eq. (\ref{e355}))
under the assumption 
that $n(\mathbf{r})$ and its gradients vanish as
$r\rightarrow\infty$. A full expression for $\tau^{(4)}$,
involving derivatives of the
density of higher than second order,
 is given in Ref.\cite{BJC}. 
Although the integration by parts that
leads to Eq. (\ref{ef4}) is inappropriate for 
some non-analytic densities \cite{PSHP}, we shall
incorporate Eq. (\ref{ef4}) into our MGGA.  
The sixth-order term \cite{Mu}, even if it 
provides a useful 
correction to the fourth-order gradient expansion
for the formation energy of a monovacancy in jellium \cite{YPKZ}, 
diverges for atoms after the integration of Eq. (\ref{e355}), 
and requires higher derivatives of the density
than we would like to use.
For later use, we define
\begin{equation}
F^{GE4}_s(p,q) = 1+(5/27)p+(20/9)q+\Delta.
\label{rlbfsge4}
\end{equation}

The von Weizs\"{a}cker expression \cite{vW} ($\tau^W=|\nabla n|^2/(8n)$)
is exact
for any one- or two-electron ground state, 
is accurate in nearly iso-orbital
regions, satisfies the exact nuclear cusp condition 
($\tau(0)=Z^2 n(0)/2$ where $Z$ is the nuclear charge) \cite{Ka},
is exact in the $r\rightarrow\infty$ asymptotic region 
(where the density matrix behaves like Eq. (8) of Ref.\cite{EBP}), and 
gives a rigorous lower bound \cite{HO1,HO2}
\begin{equation}
\tau^{W}(\mathbf{r})\leq\tau(\mathbf{r}),
\label{rlb}
\end{equation}
or $F^{W}_{s}\leq F_s$.
The semi-local bound of Eq. (\ref{rlb}) is one of the
most important constraints in the construction of our functional.
The von Weizs\"{a}cker enhancement factor \cite{vW}
\begin{equation}
F^{W}_{s}=(5/3)p\geq 0
\label{vWef}
\end{equation}
is simple. But the von  Weizs\"{a}cker KE functional gives, in general, 
very poor approximate atomization kinetic energies (see Table III of
Ref.\cite{IEMS}),
and this has been attributed \cite{We1} 
to its strong violation of Eq. (8) of Ref. \cite{We1}.

Recently Tao, Perdew, Staroverov and Scuseria (TPSS) \cite{TPSS1} have 
constructed a nonempirical meta-generalized gradient
approximation for the exchange-correlation energy.
This functional, which satisfies as many exact 
constraints as a meta-GGA can (see Table 1
of Ref.\cite{TPSS2}, or Ref.\cite{PTSS2} for a detailed
explanation), provides a good overall
description of atoms, molecules, solids and surfaces \cite{TPSS2}-\cite{CPT}. 
We construct a Laplacian-level TPSS (LL-TPSS) by replacing 
$\tau$ by $\tau^{MGGA}$ in the TPSS exchange-correlation energy per particle.

Laplacian-level functionals for exchange and correlation have been
advocated in Ref. \cite{Uk1}, and proposed in Ref. \cite{Kk1}. 
Recent interest in them is driven in
part by the observation of Ref. \cite{Ck1} that $\nabla^2 n$ 
can be used along with $n$
to mimic some exact exchange-correlation energy densities.

The paper is organized as follows. In section \ref{sec2}, 
we present the KED meta-GGA functional.
In section \ref{sec3}, we test our functional for 
several physical systems and models, and further explain its behavior.
In section \ref{sec4}, we construct the LL-TPSS
exchange-correlation functional, and present numerical and analytic evidence
that $\nabla^2 n$ and $\tau$ carry essentially 
the same information beyond that carried by $n$ and $\nabla n$.
In section \ref{sec5}, we summarize our 
conclusions.

\section{CONSTRUCTION OF A META-GGA KINETIC ENERGY DENSITY}
\label{sec2}

\noindent

Our meta-GGA for the KED is an interpolation between a modified gradient
expansion and the von Weizs\"{a}cker expression.  
There are many ways to satisfy the inequality of Eq. (\ref{rlb}), 
so we will have to rely on empiricism to select one of them.

First, we construct a modified fourth-order gradient expansion enhancement factor
\begin{equation}
F^{GE4-M}_{s}=F^{GE4}_s/\sqrt{1+(\frac{\Delta}{1+(5/3)p})^2}\;\;,
\label{FSGE4}
\end{equation}
which has the following properties: 

(1) For small $p$ and $|q|$,
\begin{equation}
F^{GE4-M}_{s}=F^{GE4}_s + O(\Delta^2),
\label{pr1}
\end{equation}
so that it recovers the fourth-order 
gradient expansion for a slowly-varying density. 

(2) When $|q|\rightarrow\infty$,
\begin{equation}
F^{GE4-M}_{s}\longrightarrow 1+F_{s}^{W},
\label{pr2}
\end{equation}
which is the correct limit for a uniform density perturbed by a small-amplitude,
short-wavelength density wave \cite{JY}. When $p\rightarrow\infty$,
$F^{GE4-M}_{s}\rightarrow F_{s}^{W}+o(p^0)$, 
which is reasonable for other rapid density variations.

(3) The modified enhancement factor of Eq. (\ref{FSGE4}) satisfies a uniform 
damping property:
\begin{equation}
|F^{GE4-M}_{s}| < |F^{GE4}_{s}|.
\label{pr3}
\end{equation}
This is desirable because for large $p$ and $|q|$ spuriously large values of
$|F^{GE4}_{s}|$ can arise from truncation of the gradient expansion.

\begin{figure}
\includegraphics[width=\columnwidth]{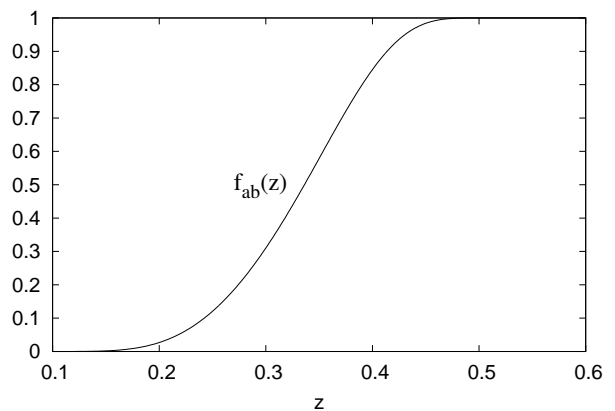}
\caption{ Interpolating function $f_{ab}(z)$ (see Eq. (\ref{pr4})) versus $z$, 
for the optimized parameters $a=0.5389$ and $b=3$.} 
\label{functionG}
\end{figure}

We shall assume that the condition $F^{GE4-M}_{s}<\sim F_{s}^ {W}$
 indicates the need for $F^{MGGA}_{s}=F_{s}^ {W}$.
Our meta-GGA interpolates between $F^{GE4-M}_{s}$ and $F_{s}^ {W}$. 
The smooth interpolating function is
\begin{equation}
f_{ab}(z)=\left\{ \begin{array}{lll}
0,     & z\leq 0\\
(\frac{1+e^{a/(a-z)}}{e^{a/z}+e^{a/(a-z)}})^b,   & 0< z< a\\
1,     & z\geq a,\\
                                    \end{array}
\right.      
\label{pr4}
\end{equation}
where $0< a\leq 1$ and $b>0$ are parameters. 
This function is plotted in Fig. \ref{functionG}.
The meta-GGA KED is defined as follows:
\begin{equation}
\tau^{MGGA}=\tau^{(0)}F_{s}^{MGGA}(p,q),
\label{pr5}
\end{equation}
where
\begin{eqnarray}
& F_{s}^{MGGA}=F_{s}^{W}+(F_{s}^{GE4-M}-F_{s}^{W})\nonumber\\
& \times f_{ab}(F_{s}^{GE4-M}-F_{s}^{W}).
\label{pr6}
\end{eqnarray}
When $f_{ab}=1$, i.e., when $F_{s}^{GE4-M}>F_{s}^{W}+a$, Eq. (\ref{pr6}) makes
$F_{s}^{MGGA}=F_{s}^{GE4-M}$.  When $f_{ab}=0$, i.e., 
when $F_{s}^{GE4-M}<F_{s}^{W}$, Eq. (\ref{pr6}) makes
$F_{s}^{MGGA}=F_{s}^{W}$. 
In between, $F_{s}^{MGGA}$ is an interpolation between $F_{s}^{GE4-M}$
and $F_{s}^{W}$.
Our positive meta-GGA KED keeps all the correct features of $\tau^{GE4-M}$, 
tends to be exact in iso-orbital regions,
and satisfies the important constraint of Eq. (\ref{rlb}):
\begin{equation}
F_{s}^{MGGA}\geq F_{s}^{W}.
\label{pr7}
\end{equation}

The meta-GGA depends on the empirical parameters $a$ and $b$. These parameters 
were found numerically, by minimizing the following expression
\begin{eqnarray}
& \mathrm{Error}=\frac{1}{2}\mathrm{``m.a.r.e. atoms"}+
\frac{1}{4}\mathrm{``m.a.r.e. clusters"}\nonumber\\
& +\frac{1}{4}\mathrm{``m.a.r.e. LDM(N=8)"}, 
\label{error}
\end{eqnarray}
where ``m.a.r.e. atoms" is the mean absolute relative error 
(m.a.r.e.) of the integrated kinetic energy of 
50 atoms and ions, ``m.a.r.e. clusters" 
is the m.a.r.e. of the KE of $2e^{-}$, $8e^{-}$, $18e^{-}$, $20e^{-}$, 
$34e^{-}$, $40e^{-}$, $58e^{-}$, $92e^{-}$, and $106e^{-}$ neutral 
spherical jellium 
clusters (with bulk parameter $r_{s}=3.93$ which corresponds to 
Na) and ``m.a.r.e. LDM(N=8)" is the m.a.r.e. of the KE of 
N=8 jellium spheres for $r_s$ = 2, 4, and 6, calculated in the 
liquid drop model (LDM)
\begin{equation}
T^{LDM}_{s}=(3/10)k^{2}_{F}N+\sigma_{s}N^{2/3}4\pi r^{2}_{s},
\label{ldm}
\end{equation}
where $k_{F}=(3\pi^{2}n)^{1/3}$ is the Fermi wavevector, 
$r_{s}=(3/4\pi n)^{1/3}$ is the radius of a sphere which 
contains on average 
one electron, $n$ is the bulk density, and $\sigma_{s}$ is the surface KE.
  The "exact" LDM value is one computed with the
exact $\sigma_{s}$. 
Because the relative errors of surface kinetic energies 
are much larger than those of the atoms and spherical jellium clusters, 
we use the LDM approach for calculating the jellium surface 
KE errors; LDM gives m.a.r.e. comparable to that of 
atoms and clusters (see Table \ref{errors}). 
The densities and orbitals we use are analytic Hartree-Fock 
\cite{CR11} for atoms
and ions, and numerical Kohn-Sham for clusters 
and surfaces (with the local
density approximation for the exchange-correlation potential).

In the process of optimization, we observed that the m.a.r.e. decreases very slowly
when the parameter $b>3$ increases, 
but large values of $b$ deteriorate the KED. So, we chose 
the following set of parameters: $a=0.5389$ and $b=3$. 
As we can see in Table \ref{errors},
this set of parameters gives an accuracy close to (but better than) that 
of the fourth-order gradient expansion.
\begin{table}[htbp]
\footnotesize
\caption{ Mean absolute relative error (m.a.r.e.) of 
integrated kinetic energies of 50 atoms and ions, of neutral spherical
jellium Na clusters ($2e^{-}$, $8e^{-}$, $18e^{-}$, 
$20e^{-}$, $34e^{-}$, $40e^{-}$, $58e^{-}$, $92e^{-}$, and $106e^{-}$)
and of jellium surfaces (with $r_{s}=2$, $r_{s}=4$, and $r_{s}=6$) 
incorporated into the liquid drop model (LDM) for a 
jellium sphere with N=8 electrons
(see Eq. (\ref{ldm})).
 The atoms and ions are: H, He, $\mathrm{Be}^{+2}$,
Be, $\mathrm{Be}^{+1}$, Li, $\mathrm{Li}^{+1}$, Ne, $\mathrm{Ne}^{+8}$,
 $\mathrm{Ne}^{+7}$, $\mathrm{Ne}^{+6}$, Ar, $\mathrm{Ar}^{+16}$, 
$\mathrm{Ar}^{+15}$, $\mathrm{Ar}^{+14}$, $\mathrm{Ar}^{+8}$, $\mathrm{Ar}^{+6}$,
Zn, $\mathrm{Zn}^{+28}$, $\mathrm{Zn}^{+20}$, $\mathrm{Zn}^{+18}$,
 $\mathrm{Zn}^{+12}$, Kr, $\mathrm{Kr}^{+34}$, $\mathrm{Kr}^{+26}$, 
$\mathrm{Kr}^{+24}$, $\mathrm{Kr}^{+18}$, Xe, $\mathrm{C}^{+4}$,
 $\mathrm{C}^{+3}$, $\mathrm{C}^{+2}$, $\mathrm{N}^{+5}$, $\mathrm{N}^{+4}$,
 $\mathrm{N}^{+3}$, $\mathrm{B}^{+1}$, $\mathrm{B}^{+3}$, $\mathrm{B}^{+2}$,
$\mathrm{O}^{+1}$, $\mathrm{O}^{+6}$, $\mathrm{O}^{+5}$, $\mathrm{O}^{+4}$,
Cu, $\mathrm{Cu}^{+27}$, $\mathrm{Cu}^{+26}$, $\mathrm{Cu}^{+25}$, 
$\mathrm{Cu}^{+19}$, $\mathrm{Cu}^{+17}$, $\mathrm{Cu}^{+11}$,
$\mathrm{Cu}^{+1}$ and N. }
\begin{tabular}{|l|l|l|l|l|l|}
   \multicolumn{1}{c}{ } &
   \multicolumn{1}{c}{ } &
   \multicolumn{1}{c}{ } &
   \multicolumn{1}{c}{ } &
   \multicolumn{1}{c}{ }\\  \hline
 & $T^{(0)}_{s}$ & $T^{(0)}_{s}+T^{(2)}_{s}$ & $T^{(0)}_{s}+T^{(2)}_{s}+T^{(4)}_{s}$ & 
$T^{MGGA}_{s}$ \\  \hline
m.a.r.e. atoms & 0.0842 & 0.0112 & 0.0251 & 0.0139   \\  \hline
m.a.r.e. clusters & 0.0439 & 0.0099 & 0.0176 & 0.0245  \\  \hline
m.a.r.e. LDM(N=8) & 0.0810 & 0.0330 & 0.0170 & 0.0247 \\  \hline
Error (Eq. (\ref{error})) & 0.0733 & 0.0163 & 0.0212 & 0.0193 \\  \hline
\end{tabular}
\label{errors}
\end{table}
%
%
%

In Fig. \ref{functionG}, we plot the 
interpolating function $f_{ab}(z)$ using our choice 
for the parameters.
Near a nucleus, there is a large region where $F^{GE4}_{s}<0$
and thus $F^{GE4-M}_{s}<0$,
making $f_{ab}=0$ and $F^{MGGA}_{s}=F^{W}_{s}$,
the correct behavior.
(Nevertheless, as $r\rightarrow 0$ and $q\rightarrow -\infty$, 
$F^{GE4-M}_{s}\rightarrow 1+F^{W}_{s}$, making 
$f_{ab}\rightarrow 1$
and $F^{MGGA}_{s}\rightarrow 1+F^{W}_{s}\approx 1.24$, 
which is at least positive and finite.)  In the 
$r\rightarrow\infty$ asymptotic region, where 
$p\approx q\rightarrow\infty$, $F^{GE4-M}_{s}\rightarrow F^{W}_{s}$, 
making
$f_{ab}\rightarrow 0$ and $F^{MGGA}_{s}\rightarrow F^{W}_{s}$, 
the correct limit.
In the slowly-varying limit, where $p$ and $q$ 
are small, $f_{ab}(z)\rightarrow 1$ , so meta-GGA recovers the 
fourth-order 
gradient expansion.
%
\begin{figure}
\includegraphics[width=\columnwidth]{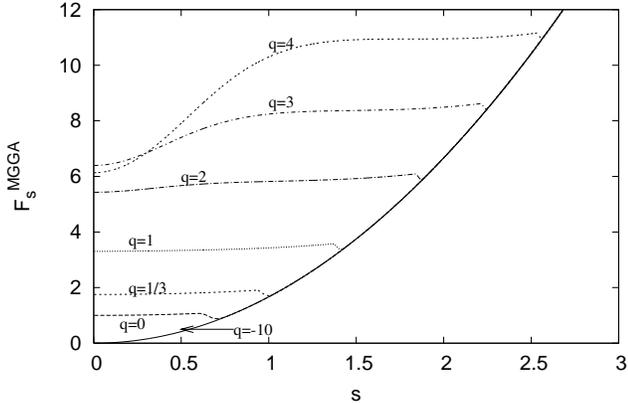}
\caption{ Enhancement factor $F^{MGGA}_{s}(p,q)$ versus 
reduced gradient $s=\sqrt{p}$, 
for several values of the reduced Laplacian 
$q$ (-10, 0, 1/3, 1, 2, 3, and 
4). The parabolic asymptote is $F^W_s$.}
\label{Efactor}
\end{figure}
In Fig. \ref{Efactor}, we plot the meta-GGA enhancement 
factor versus $s=\sqrt{p}$, for different values of $q$. 
We observe an orderly behavior of $F^{MGGA}_{s}$ for $p$ and $q$ 
in the range appropriate to physical densities.
For the integrated KE, our meta-GGA 
is size-consistent, and satisfies the uniform scaling relation 
\cite{LP}
and the spin-scaling relation \cite{a57}.
The meta-GGA nonuniform scaling behavior is investigated in Section \ref{sec3}.

\section{RESULTS: KINETIC ENERGY AND ITS DENSITY}
\label{sec3}
\noindent

\begin{figure}
\includegraphics[width=\columnwidth]{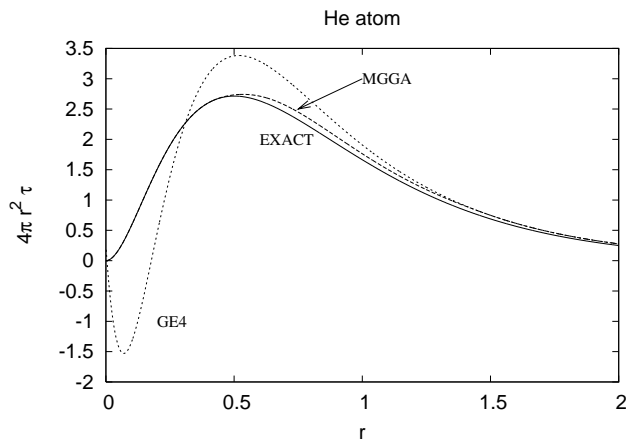}
\caption{ KE integrand $4\pi r^{2}\tau$ versus radial distance r for the He
atom. 
The integral under the curve is the KE for the helium atom: exact: 2.862 a.u., 
meta-GGA: 2.993 a.u., and GE4: 2.963 a.u..}
\label{Heatom}
\end{figure}

\begin{figure}
\includegraphics[width=\columnwidth]{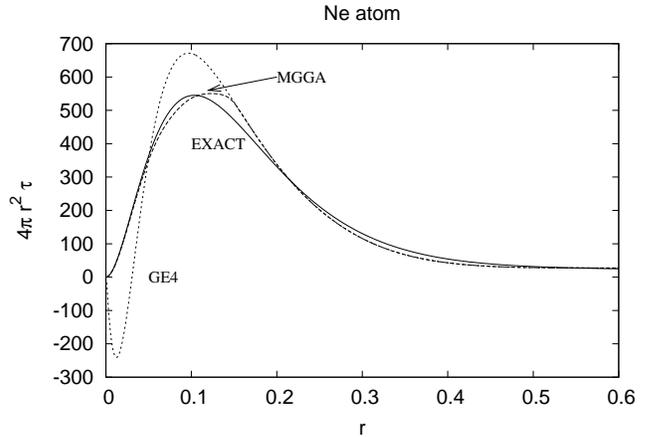}
\caption{ KE integrand $4\pi r^{2}\tau$ versus 
radial distance r for the Ne atom. 
The integral under the curve is the KE for the neon atom:
exact: 128.546 a.u.,
meta-GGA: 129.312 a.u., and GE4: 129.749 a.u..}
\label{Neatom}
\end{figure}

In Figs. \ref{Heatom} and \ref{Neatom} we plot the integrand of
the KE, $4\pi r^{2}\tau$, versus radial distance 
from the nucleus, for the He and Ne atoms. The KED of the 
fourth-order gradient
expansion, $\tau^{GE4}$, 
is negative near the nucleus and is not a good approximation for 
$\tau$. 
The meta-GGA 
KE integrand is much improved near the nucleus, and
everywhere follows
very nicely the exact behavior. 
%

\begin{figure}
\includegraphics[width=\columnwidth]{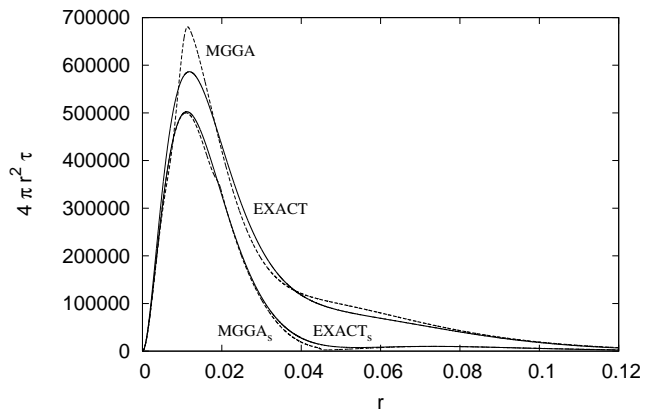}
\caption{ KE integrand $4\pi r^{2}\tau$ versus radial distance 
r for the 
10-electron (hydrogenic orbitals) ion with nuclear charge Z=92. 
The curves $\mathrm{EXACT}_s$ and $\mathrm{MGGA}_s$ show the 
contribution of the four s-electrons (see Ref.\cite{HMD}).}
\label{shell}
\end{figure}

Let us consider an ion model with 10 electrons which occupy the 
first hydrogenic orbitals and with a nuclear charge Z=92. In such a 
closed-shell 
system, the KED is determined by the density of the 
$s$-electrons alone \cite{HMD,MS}. In Fig. \ref{shell}, we show the 
exact and the meta-GGA kinetic energy densities for this system
(for the s-electrons, and for the whole system, separately). 
%

\begin{figure}
\includegraphics[width=\columnwidth]{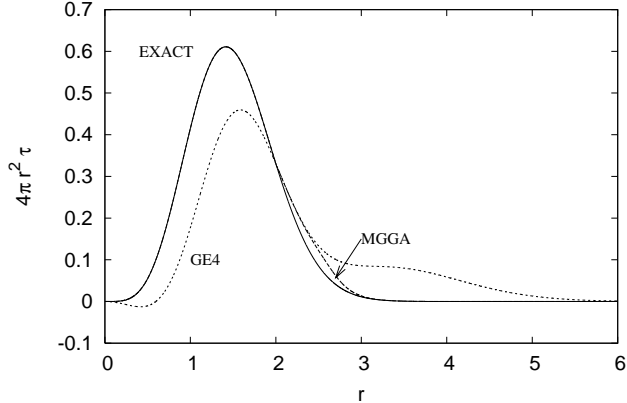}
\caption{ KE integrand $4\pi r^{2}\tau$ versus radial distance 
r for
the one-electron Gaussian density.
The integral under the curve is the kinetic energy: $\mathrm{exact}=0.750$ a.u.,
$\mathrm{meta}$-$\mathrm{GGA}=0.778$ a.u., and $\mathrm{GE4}=0.865$ a.u. }
\label{GAUSS}
\end{figure}

The KED of the one-electron Gaussian density is shown in Fig. \ref{GAUSS}.
This system does not have a cusp near the center ($r=0$, where $p=0$ and
$q\approx -0.5$) and, in this sense, it is an important test for our meta-GGA,
which we find to be as accurate here as it is for the He atom.
%
%

\begin{figure}
\includegraphics[width=\columnwidth]{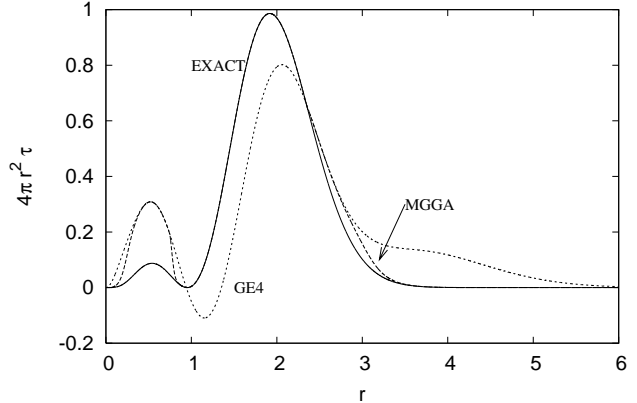}
\caption{ KE integrand $4\pi r^{2}\tau$ versus 
radial distance r for 
the pseudo-Hooke's atom in the high-correlation case, using the scaled density 
given in Eq. (\ref{hooke2}). 
The integral under the curve is the kinetic energy: $\mathrm{exact}=1.115$ a.u.,
$\mathrm{meta}$-$\mathrm{GGA}=1.264$ a.u., and $\mathrm{GE4}=1.185$ a.u. }
\label{Hooke}
\end{figure}

The Hooke's atom is a simple system of two interacting electrons 
in a harmonic potential. For this system, the exact KED is 
the von Weizs\"{a}cker one. 
The correlated wavefunction and its density can 
be calculated exactly \cite{Taut} for special values of
the spring constant $k$.  The low-correlation case \cite{FUT}
corresponds to $k = 0.25$a.u., 
and the high-correlation case \cite{FGU} 
to  $k=3.6\times 10^{-6}$ a.u.. 
A modeled density very similar to the exact Hooke's 
atom density (with $k=3.6\times 10^{-6}$ a.u.) is 
\begin{equation}
n(r)=A(1+B\;C\;r^{2})e^{-Cr^{2}},
\label{hook1}
\end{equation}   
where $A=0.67\times 10^{-6}$ a.u., B=11 a.u., and 
C=0.0010212 a.u. Applying uniform scaling $n(\mathbf{r})\rightarrow 
\gamma^{3}n(\gamma 
\mathbf{r})=n_{\gamma}(\mathbf{r})$, we define the following scaled density:  
\begin{equation}
n_{\gamma}(r)=0.02145(1+10.5r^{2})e^{-r^{2}},
\label{hooke2}
\end{equation} 
with $\gamma=31.753$.
In Fig. \ref{Hooke} we show the exact, meta-GGA and fourth-order 
gradient expansion KE integrands for the pseudo-Hooke's atom 
(using the scaled density given in Eq. (\ref{hooke2})) in
the high-correlation case. The region near the nucleus of 
the pseudo-Hooke's atom is unusual because there
strong correlation creates a deep ``hole" in 
the density: $q$ decreases smoothly
with increasing $r$ (from 19.29 at $r=0$ to
$\approx 0$ at $r=0.68$), and $p$ increases slowly 
with increasing $r$ (from 0 at $r=0$ to a peak value of 1.829 at $r=0.233$).
This region can not be described accurately 
by the meta-GGA, as we can see in Fig. \ref{Hooke}. 
Near $r=1$, where $p\simeq 0$ 
and $q\simeq -0.495$, the KED meta-GGA  recovers the exact behavior. 
%
%

\begin{figure}
\includegraphics[width=\columnwidth]{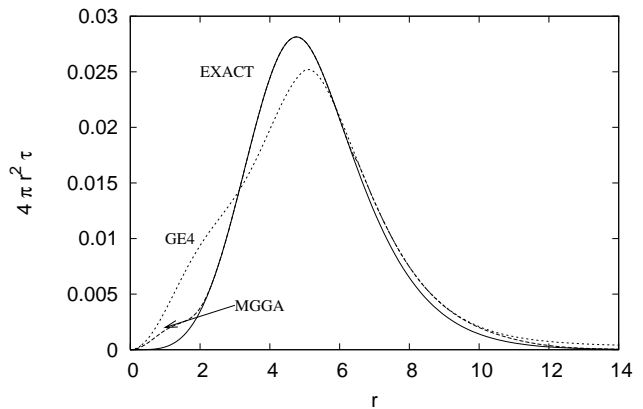}
\caption{ KE integrand $4\pi r^{2}\tau$ versus 
radial distance r for 
the $2e^{-}$ Na jellium spherical cluster. 
The integral under the curve is the kinetic 
energy:
$\mathrm{exact}=0.114$ a.u.,
$\mathrm{meta}$-$\mathrm{GGA}=0.120$ a.u., and $\mathrm{GE4}=0.124$ a.u. }
\label{cluster}
\end{figure}

Figs. \ref{Hooke} and \ref{cluster} show how our 
$\tau^{MGGA}$ can sometimes fail to
recognize iso-orbital regions where $\tau=\tau^W$.  
For the $2e^{-}$ jellium cluster between $r=0.1$ and $r=1$, where 
$p$ increases very slowly with increasing $r$ (from $p=0.0016$ at $r=0.1$
to $p=0.015$ at $r=1$) and 
$q\simeq -0.29$, the meta-GGA functional switches  
from the fourth-order 
gradient expansion to the exact behavior due to the construction 
of the interpolating function $f_{ab}$. 
This feature is important for molecules, 
because a similar case arises at the center of a 
diatomic molecule $X_2$.  
%
%

\begin{figure}
\includegraphics[width=\columnwidth]{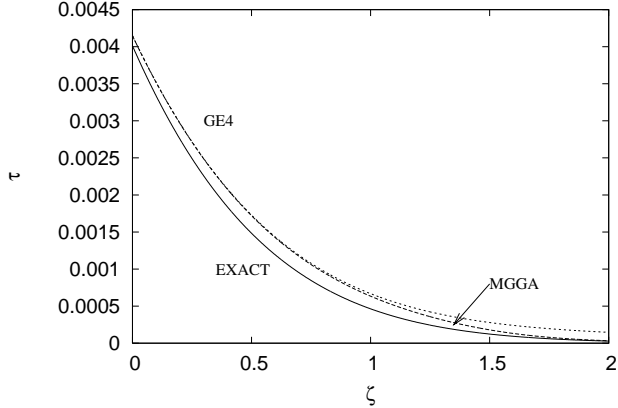}
\caption{ Kinetic energy density versus Airy 
scaled distance \cite{VJKS} for the Airy gas. The exact KED 
is given by Eq. (11) of Ref.\cite{VJKS}.}
\label{Airy2}
\end{figure}

In Fig. \ref{Airy2} we plot the KED $\tau$ 
versus the scaled distance $\zeta$ for the Airy gas, 
which is 
the simplest model of an edge electron gas \cite{KM}. 
In this model, the non-interacting electrons see
a linear potential.  In the tail of the density, 
the KED meta-GGA becomes exact (see
Fig. \ref{Airy2}) as we mentioned in Sec. \ref{sec2}.
At least two
orbital-free kinetic energy functionals have been based upon the Airy gas model.
In Ref. \cite{VJKS}, the kinetic energy density 
of the Airy gas is transferred to other
systems in a Local Airy Gas (LAG) approximation, which seems accurate for
jellium surfaces but makes $\tau$ diverge at nuclei.  
In Ref. \cite{Ba}, a density
functional is constructed for the linear potential; it has an unphysically
rapid oscillation \cite{DG}
in its correction to $\tau^{(0)}$ for a slowly-varying density.

%
%

\begin{figure}
\includegraphics[width=\columnwidth]{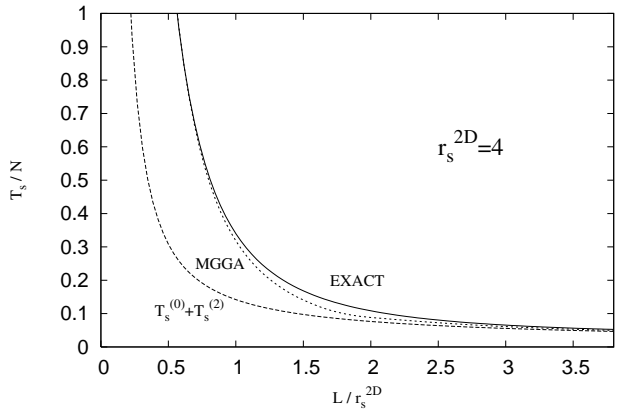}
\caption{Integrated kinetic energy per electron 
versus $L/r^{2D}_{s}$ for a quasi-2D electron gas
in the infinite barrier model. 
The exact curve is given by Eq. (\ref{ibm2d}). 
$r_s^{2D}$ is the areal density  parameter of 
the quasi-2D electron gas and L is the width of the 
quantum well. ($L< 3.85r_s^{2D}$, see Ref.\cite{PP}) 
Note that $T^{(4)}_s$ diverges here.  }
\label{Lisa}
\end{figure}

For a quasi-2D electron gas (quantum well) whose orbitals are
those of noninteracting electrons confined by 
infinite barriers \cite{PP}, the exact
KE per electron can be calculated analytically, and has the simple expression:
\begin{equation}
\frac{T_s}{N}=\frac{\pi^2}{2L^2}+\frac{(k_{F}^{2D})^2}{4},
\label{ibm2d}
\end{equation}
where L is the width of the quantum well, 
$k_{F}^{2D}=\sqrt{2}/r_s^{2D}$ is the two-dimensional 
Fermi wavevector and
$r_s^{2D}$ is the radius of the circle that contains on average one electron of 
the quasi two-dimensional gas. 
 Fig. \ref{Lisa} is for the electron gas at
$r_s^{2D} = 4$.  $T^{(4)}_{s}$ diverges due to the infinite barriers, 
so it is not shown in the figure. As we can see, the KED
meta-GGA functional solves this nonuniform scaling
problem almost exactly.
%
%

\begin{figure}
\includegraphics[width=\columnwidth]{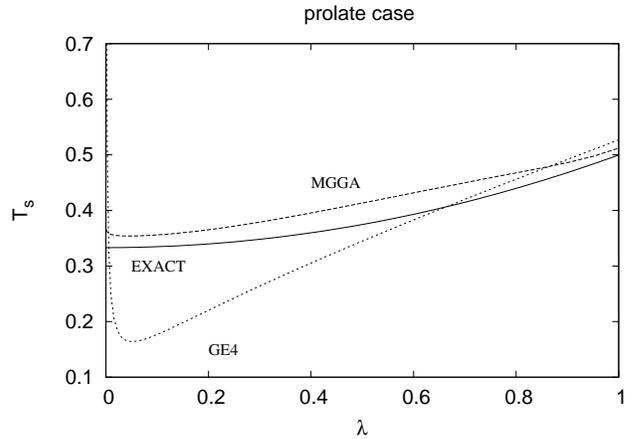}
\caption{ Integrated KE versus scaling parameter $\lambda$ for the 
nonuniformly-scaled hydrogen atom, in the prolate case. }
\label{Hprolate}
\end{figure}

\begin{figure}
\includegraphics[width=\columnwidth]{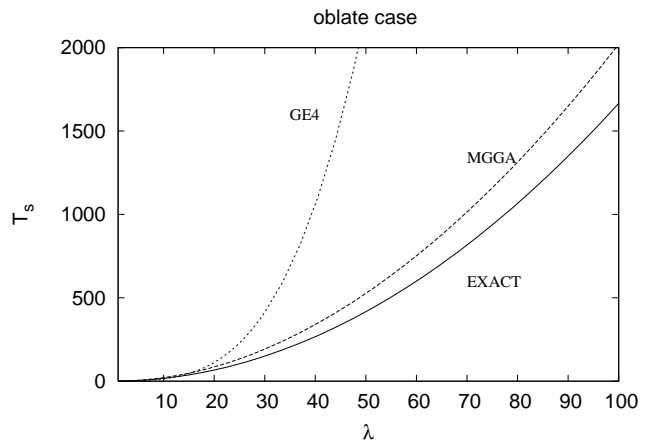}
\caption{Integrated KE versus scaling parameter $\lambda$ 
for the nonuniformly-scaled hydrogen atom, in the oblate case. }
\label{Hoblate}
\end{figure}

The nonuniform scaling inequality \cite{YL} is 
\begin{equation}
T_{s}[n^{x}_{\lambda}]\leq\lambda^{2}T^{x}_{s}[n]+T^{y}_{s}[n]+T^{z}_{s}[n],
\label{nus}
\end{equation}
where $\lambda$ is a positive scale factor, the nonuniformly-scaled 
density is 
$n^{x}_{\lambda}(x,y,z)=\lambda n(\lambda x,y,z)$, and 
$T^{q}_{s}[n]=(1/2)(\partial 
T_{s}[n^{q}_{\lambda}]/ \partial\lambda)_{\lambda=1}$, 
where $q$ is $x$, $y$ or $z$. 
For the von Weizs\"{a}cker functional, and thus for an exact
treatment of the nonuniformly-scaled hydrogen atom of density
$n_{\lambda}(\mathrm{r})=(\lambda/\pi)\exp(-2\sqrt
{(\lambda x)^{2}+y^{2}+z^{2}})$
, Eq. (\ref{nus}) becomes an equality \cite{YL}. 
In Figs. \ref{Hprolate} and \ref{Hoblate},
we show the KE  of the nonuniformly-scaled hydrogen atom as a function of
$\lambda$, for the prolate case ($\lambda\leq 1$) and
oblate case ($\lambda\geq 1$) respectively.
As we can see, the meta-GGA KE functional does not satisfy the
nonuniform-scaling inequality, but is still very close to the exact
behavior.
The von Weizs\"{a}cker KE is always less than or
equal to the meta-GGA functional (see Eq. (\ref{pr7})).
So, the meta-GGA  
seems to describe the nonuniform scaling relation with considerable 
fidelity, a potentially important feature for molecules
where the bonding causes nonuniform distortions in the density.  
%
%
%

When a molecule at its equilibrium geometry is broken up into separate
atoms, the total energy increases and (as suggested by the virial theorem)
the kinetic energy decreases.
In Table \ref{molecules} we present the atomization kinetic energies for a set of
molecules used in Ref.\cite{IEMS}. The meta-GGA kinetic energy functional gives the best 
overall results, but is still not accurate enough for chemical applications.
We observe that for $\mathrm{NO}$ and $\mathrm{O}_2$ molecules, where the other listed
functionals fail badly, the meta-GGA keeps the right sign. (In fact, with the exception 
of $\mathrm{N}_2$ and $\mathrm{F}_2$, the atomization kinetic energies of the 
meta-GGA all have the right sign.)

\begin{table}[htbp]
\footnotesize
\caption{ Integrated atomization kinetic energy 
( KE atoms - KE molecule, in a. u.) for a few 
small molecules.
The kinetic energies were calculated 
using the PROAIMV code with Kohn-Sham orbitals given by the Gaussian 2000 code
(with the uncontracted $6-311+G(3df,2p)$ basis set, Becke 1988 exchange functional
\cite{Becke},
and Perdew-Wang correlation functional \cite{PW91}).
The last line shows the mean absolute errors (m.a.e.).}
\begin{tabular}{|l|l|l|l|l|l|} 
   \multicolumn{1}{c}{ }&
   \multicolumn{1}{c}{ }&
   \multicolumn{1}{c}{ }&
   \multicolumn{1}{c}{ }&
   \multicolumn{1}{c}{ } \\  \hline
 & $T^{\mathrm{exact}}_{s}$ & $T^{(0)}_{s}$ & $T^{(0)}_{s}+T^{(2)}_{s}$ & 
$T^{(0)}_{s}+T^{(2)}_{s}+T^{(4)}_{s}$ & $T^{MGGA}_{s}$ \\  \hline
$\mathrm{H}_2$ & -0.150 & -0.097 & -0.114 & -0.119 & -0.216 \\  \hline
$\mathrm{HF}$  & -0.185 & -0.305 & -0.186 & -0.133 & -0.352 \\  \hline
$\mathrm{H}_{2}\mathrm{O}$ & -0.304 & -0.308 & -0.136 & -0.057 & -0.634 \\  \hline
$\mathrm{CH}_4$  & -0.601 & -0.737 & -0.571 & -0.498 & -1.036 \\  \hline
$\mathrm{NH}_3$  & -0.397 & -0.231 & -0.060 & 0.014 & -0.477 \\  \hline
$\mathrm{CO}$  & -0.298 & -0.323 & -0.085 & 0.015 & -0.458 \\  \hline
$\mathrm{F}_2$  & -0.053 & 0.128 & 0.282 & 0.338 & 0.154 \\  \hline
$\mathrm{HCN}$  & -0.340 & -0.1835 & 0.079 & 0.186 & -0.328 \\  \hline
$\mathrm{N}_2$  & -0.158 & 0.344 & 0.565 & 0.650 & 0.319 \\  \hline
$\mathrm{CN}$  & -0.431 & -0.215 & 0.005 & 0.094 & -0.231 \\  \hline
$\mathrm{NO}$  & -0.268 & 0.092 & 0.330 & 0.422 & -0.084 \\  \hline
$\mathrm{O}_2$  & -0.100 & 0.106 & 0.335 & 0.431 & -0.194 \\  \hline
m.a.e. &               & 0.177 & 0.311 & 0.384 & 0.201 \\  \hline
\end{tabular}
\label{molecules}
\end{table}

\section{LAPLACIAN-LEVEL META-GGA FOR EXCHANGE-CORRELATION ENERGY}
\label{sec4}
\noindent

The exchange-correlation meta-GGA uses as ingredients the spin densities and 
their gradients, and the positive Kohn-Sham kinetic energy
densities:
\begin{eqnarray}
&E^{MGGA}_{xc}=\int d\R \; n(\R)\epsilon^{(0)}_{x}(n(\R))\nonumber\\ 
&\times F^{MGGA}_{xc}(\UP(\R),\DN(\R),\nabla\UP(\R),\nabla\UP(\R),
\tau_{\uparrow}(\R),\tau_{\downarrow}(\R)),
\label{MGGA1ll}
\end{eqnarray}
where $\epsilon^{(0)}_x(n) = (-3/4\pi)(3\pi^2 n)^{1/3}$ is
the exchange energy per electron of a uniform, spin-unpolarized density n.
The local density approximation (LDA) is recovered for $\nabla \UP = 
\nabla \DN =0$ and $\tau_{\uparrow}=\tau^{(0)}_{\uparrow}$, 
$\tau_{\downarrow}=\tau^{(0)}_{\downarrow}$.
Our Laplacian-level non-interacting KED functionals
$\tau_{\uparrow}^{MGGA}(\mathbf{r})$ and 
$\tau_{\downarrow}^{MGGA}(\mathbf{r})$ 
can replace the exact Kohn-Sham kinetic energy
densities $\tau_{\uparrow}(\mathbf{r})$ and 
$\tau_{\downarrow}(\mathbf{r}))$ in Eq. (\ref{MGGA1ll}).
The resulting  LL-MGGA is an explicit density
functional, while Eq. (\ref{MGGA1ll}) is an implicit one.

We have constructed in this way the Laplacian-level (LL-TPSS) of the 
Tao, Perdew, Staroverov and 
Scuseria exchange-correlation meta-GGA \cite{TPSS1}.
Like TPSS, LL-TPSS recovers the fourth-order 
gradient expansion \cite{a103,TPSS1} for the exchange energy 
in the slowly-varying limit.
Moreover,
like TPSS, LL-TPSS has a finite exchange potential at a nucleus.  
The need for an ingredient beyond $n$ and $\nabla n$ (e.g., $\nabla^2 n$) 
to satisfy this exact
constraint was emphasized in Refs. \cite{Uk1} and \cite{FGU}.
The
nuclear cusp of an atom can be defined by $q\rightarrow -\infty$ and 
$s\approx 0.376$, so the constraint used in the construction
of the TPSS exchange enhancement factor \cite{TPSS1}
\begin{equation}
dF^{TPSS}_{x}(s,z=1)/ds|_{s=0.376}=0,
\label{q400}
\end{equation}
where $z=\tau(\R)/\tau^{W}(\R)$, becomes
\begin{equation}
dF^{LL-TPSS}_{x}(s,q\rightarrow -\infty)/ds|_{s=0.376}=0.
\label{q401}
\end{equation}
Such constraints can be satisfied 
by a Laplacian-level meta-GGA, but not by a GGA (using only 
$n$ and $\nabla n$). 

\begin{figure}
\includegraphics[width=\columnwidth]{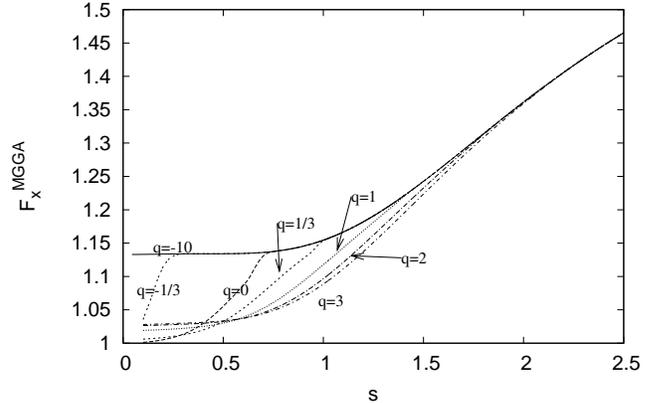}
\caption{ Exchange enhancement factor $F^{LL-TPSS}_{x}$ 
versus reduced gradient 
$s=\sqrt{p}$, for several
values of the reduced Laplacian:
$q=-10, -1/3, 0, 1/3, 1, 2, 3$. }
\label{fig1}
\end{figure}

In Fig. \ref{fig1} we show the exchange enhancement factor 
$F^{LL-TPSS}_{x}$ as a function of the inhomogeneity parameter
$s=\sqrt{p}$ for several values of the reduced Laplacian
$q$. 
The enhancement factor interpolates in an orderly
way between the exact slowly-varying limit 
(for small $s$ and $|q|$) and a rapidly-varying limit 
(for large $s$), while satisfying Eq.(\ref{q401}).
%
%

\begin{figure}
\includegraphics[width=\columnwidth]{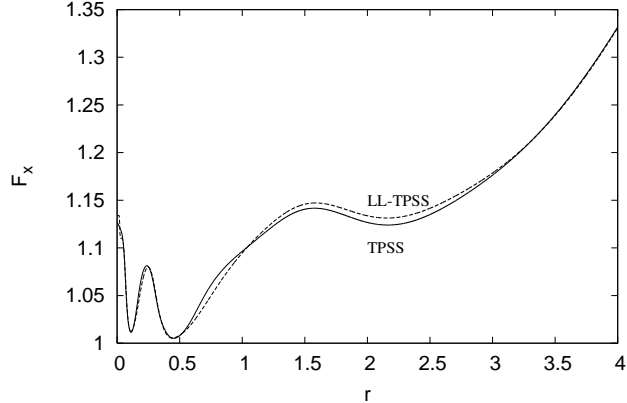}
\caption{ Exchange enhancement factors $F^{LL-TPSS}_{x}$ and 
$F^{TPSS}_{x}$ versus radial distance
r for the Zn atom. $E^{TPSS}_{x}=-69.798$ a. u.,  $E^{LL-TPSS}_{x}=-69.528$ a.u..
Note that $<r^{-1}>^{-1}$ is 0.65 for the $3d$
and 2.26 for the $4s$ electrons.}
\label{fig2}
\end{figure}

\begin{figure}
\includegraphics[width=\columnwidth]{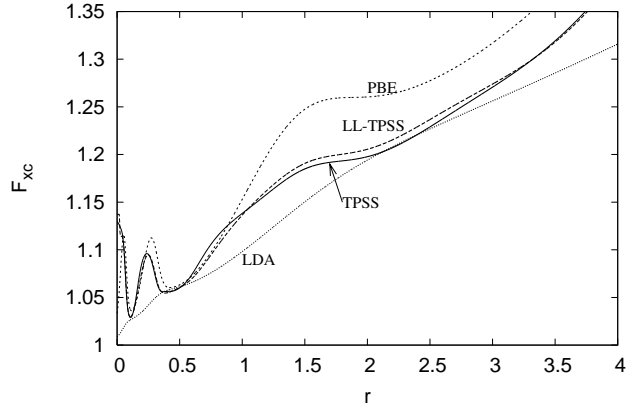}
\caption{ Exchange-correlation enhancement factors 
$F^{LL-TPSS}_{xc}$, $F^{TPSS}_{xc}$, $F^{PBE}_{xc}$ and $F^{LDA}_{xc}$ versus
radial distance
r for the Zn atom. $E^{TPSS}_{xc}=-71.208$ a.u., 
$E^{LL-TPSS}_{xc}=-71.073$ a.u.,$E^{PBE}_{xc}=-70.934$ a.u..
Note that $<r^{-1}>^{-1}$ is 0.65 for the $3d$
and 2.26 for the $4s$ electrons. PBE is the non-empirical GGA of 
Ref.\cite{PBE}}
\label{fig4}
\end{figure}

In Figs. \ref{fig2} and \ref{fig4} we compare the TPSS 
and the LL-TPSS enhancement factors (for exchange and 
exchange-correlation) for the Zn atom, which was also studied
in Ref. \cite{PTSS2}. 
The LL-TPSS exchange-correlation
 energies are close to the TPSS values for this and other atoms. 
(For the H atom, $E^{TPSS}_c=0$ and $E^{LL-TPSS}_c=-1.537\cdot 10^{-6}$;
for the other 49 atoms and ions, the m.a.r.e. of LL-TPSS with respect to
TPSS is 0.00132 for exchange, 
0.0098 for correlation, and 0.00133 for the combined 
exchange-correlation energy.) 
%
%
%
%

\begin{figure}
\includegraphics[width=\columnwidth]{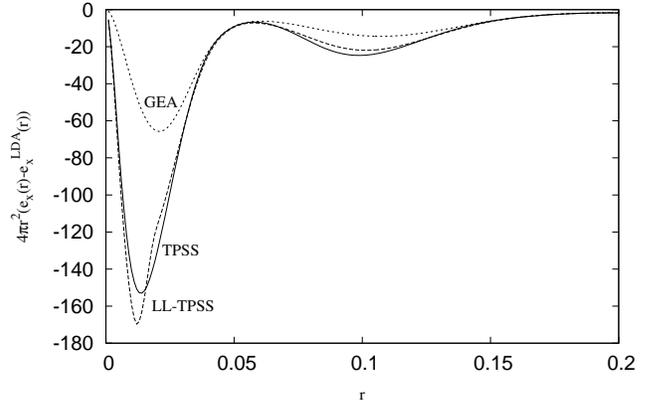}
\caption{ Deviation (from local density approximation) 
of the exchange energy 
integrand $4\pi r^2 e_x$ versus radial distance r for the Xe atom.
GEA is the second-order gradient expansion for exchange.
$E^{TPSS}_{x}=-178.449$ a.u. and $E^{LL-TPSS}_{x}=-178.240$ a.u.. }
\label{fig5}
\end{figure}

In Fig. \ref{fig5} we show that the LL-TPSS exchange
functional, like the TPSS, shows a strong enhancement 
in the $1s$ region of an atom but is elsewhere not so 
different from the second-order gradient
expansion for exchange, as discussed in Ref. \cite{PCSB}.
%

\section{Conclusions}
\label{sec5}
\noindent

We have constructed a Laplacian-level 
kinetic energy density (KED) meta-GGA functional which depends on two 
empirical parameters. These parameters control
an interpolation between a modified fourth-order 
gradient expansion and the von Weizs\"{a}cker expression.  
This interpolation is designed to respect the exact
constraint of Eq. (\ref{rlb}).  
From our tests and results 
of section \ref{sec3}, we conclude 
that the KED meta-GGA functional is a successful approximation to the 
positive Kohn-Sham KED. 

Our functional uses a simplified expression for the fourth-order gradient 
expansion \cite{Ho} of $\tau$. We also tested the meta-GGA with the full fourth-order 
terms \cite{BJC}, but did not find a considerable improvement in KED 
or integrated KE. 

We have built a Laplacian-level meta-GGA 
for exchange and correlation (LL-TPSS)
which seems to imitate faithfully the TPSS \cite{TPSS1}
exchange-correlation meta-GGA.
Several constraints of the TPSS meta-GGA are exactly
satisfied by the LL-TPSS, but others, like $E_{c}=0$ for any one-electron
system, are approximately satisfied.
It appears to us that $\nabla^2 n$ and $\tau$ carry
essentially the same information beyond that carried 
by $n$ and $\nabla n$.  Either
$\nabla^2 n$ or $\tau$ can be and are used to 
recover the fourth-order gradient
expansion of the exchange energy in the 
slowly-varying limit
 and to make the
exchange potential finite at a nucleus.

Integration of our Laplacian-level meta-GGA for the kinetic energy density
yields an orbital-free density functional for the kinetic energy that seems to
improve upon the fourth-order gradient expansion, especially for rapidly-
varying densities.  We have made all our tests for electron densities
constructed from orbitals, and do not know what might be found from a
selfconsistent solution of the orbital-free Euler equation for the electron
density.   It was
argued in Ref. \cite{WSB} that KE functionals employing 
only $n$ and $\nabla n$ cannot
yield both accurate integrated energies and accurate functional derivatives.
We believe that accurate results could be expected in some rapidly-varying
regions (near nuclei and in density tails) from KE functionals that employ $n$,
$\nabla n$, and $\nabla^2 n$; our specific expressions however may encounter a
problem due to the sharper features in Fig. \ref{Efactor} 
(also visible in Fig. \ref{fig1}).
It is generally believed that correct
quantum density oscillations and shell-structure oscillations can only be found
from a fully-nonlocal orbital-free density functional for the kinetic energy
\cite{WC,BC11}.

\begin{acknowledgment}
The authors thank L.M. Almeida for help with the surface code,
 and Lisa Pollack for early tests like those of Fig. \ref{Lisa}.
This work was supported in part by the National Science Foundation
under Grants No. DMR 01-35678 and No. DMR 05-01588.
\end{acknowledgment}  

*Present address: Donostia International Physics Center (DIPC),
Donostia, Basque Country, Spain.

\end{document}